Slopaganda: The interaction between propaganda and generative AI

Michał Klincewicz, Tilburg University
Mark Alfano, Macquarie University
Amir Ebrahimi Fard, Independent Researcher

## Abstract

At least since Francis Bacon, the slogan "knowledge is power" has been used to capture the relationship between decision-making at a group level and information. We know that being able to shape the informational environment for a group is a way to shape their decisions; it is essentially a way to make decisions for them. This paper focuses on strategies that are intentionally, *by design*, impactful on the decision-making capacities of groups, effectively shaping their ability to take advantage of information in their environment. Among these, the best known are political rhetoric, propaganda, and misinformation. The phenomenon this paper brings out from these is a relatively new strategy, which we call *slopaganda*. According to *The Guardian*, News Corp Australia is currently churning out 3000 "local" generative AI (GAI) stories each week. In the coming years, such "generative AI slop" will present multiple knowledge-related (epistemic) challenges. We draw on contemporary research in cognitive science and artificial intelligence to diagnose the problem of slopaganda, describe some recent troubling cases, then suggest several interventions that may help to counter slopaganda.

**Introduction**

From the perspective of cognitive science, individual decision-making is not a moment of endogenous choice, marked by absolute freedom. Instead, it is a series of exploratory, complex, and multi-stage processes that culminate in the execution of some action and simultaneous prediction about its consequences (Haggard 2008). Particular decisions use accumulated information about the consequences of actions at every prior stage of this process. If the information that decision-making mechanisms use is correct or adaptive, it often leads to an optimal decision and predicted consequences. We then have a robust sense of agency associated with our actions. If information is inaccurate or maladaptive, it may lead to unpredictable and undesirable consequences and a diminished sense of agency. In both cases, however, a decision leads to further accumulation of information about consequences. If the prediction was correct, there is reinforcement. In case the decision led to undesirable or unpredicted consequences, including maladaptive behavior, we should, all things considered, get revisions to the accumulated knowledge that informs future decisions.

There is a further important complication to this idealized picture of human decision-making: the world outside of the laboratory can be informationally hostile, meaning that some of the information agents can use to update their internal models of the world is maladaptive or incorrect *by design* (Sterelny 2003; Timms & Spurrett 2023). On an individual level, most people know that it is not always advantageous to foster justified and true beliefs in others. Deceptions, lies, and simply passing things over in silence are all ways in which agents can gain an epistemic and through that a practical advantage over one another's decision-making. On a group level, incorrect or maladaptive information can influence elections, shape institutional policy, and even start wars. At least since Francis Bacon, the slogan "knowledge is power" has been used to capture this relationship between decision-making at a group level and information. Shaping the informational environment for a group is a way to indirectly shape their decisions.

There are controversies about what counts as a strategy to achieve that end and some are better known than others. Karl Marx, for example, claimed that "religion is the opiate of the masses," effectively bamboozling them into submission to political and economic oppression. Foucault famously coined the phrase "power-knowledge" (*le savoir-pouvoir*) to characterize the connection between the epistemic and the practical (Foucault 2019). We focus on uncontroversial cases: strategies that *by design* influence the decision-making capacities of groups, effectively shaping their ability to take advantage of information in their environment. Among these, the best known are political rhetoric, propaganda, and misinformation (Benkler et al. 2018). We then introduce a relatively new strategy, which we call *slopaganda*. According to *The Guardian*, News Corp Australia is currently churning out 3000 "local" generative AI (GAI) stories each week. In the coming years, such "generative AI slop," as it has been dubbed in *NY Magazine*, will present multiple knowledge-related (epistemic) challenges.[1] This is not mere speculation: early results suggest that GAI is better at producing persuasive disinformation than humans are (Spitale et al. 2023). The analysis uncovers a complex interaction between the psychological mechanisms of decision-making and the informational environment that agents and groups find themselves in today. As the term 'slopaganda' is our neologism, we cannot offer a conceptual analysis that could be tested against ordinary language and intuition. As we use the term, it refers to a combination of a relatively familiar phenomenon (propaganda) and a relatively recent but prominent one (generative AI slop). Following Roberts et al. (2018; see also Stanley 2015), we understand propaganda to be the intentional manipulation of beliefs to achieve political ends. 'Slop' was only coined last year to refer to unwanted AI-generated content. Putting these together, we understand slopaganda to be unwanted AI-generated content that is spread in order to manipulate beliefs to achieve political ends.

---

[1] As far as we know, the original usage of this coinage was in url = < https://simonwillison.net/2024/May/8/slop >, accessed 15 January 2025.

Slopaganda may serve multiple purposes, including malfeasance. Following Morton (2004), we think that it is useful to distinguish three levels of social organization that are typically involved with malfeasance and atrocity at scale. This distinction will also shed light on the purposes for which slopaganda can be used. At the lowest level, there are the foot soldiers of a social movement. These are the people who *do* things, such as the brownshirts of the nazi party. At that level, information shaping strategies, including slopaganda, offer a rationale for the *doing*, especially if it involves violence, and may even persuade others to join the foot soldiers by justifying claims or up-till-then tacitly held beliefs. These are the large-scale consumers and perhaps primary targets of propaganda and perhaps also slopaganda. Next, there is the middle tier of bureaucrats — the sort of people that Hannah Arendt discusses in *Eichmann in Jerusalem* (1963). These are the people who *authorize* things. Slopaganda, like propaganda, gives reasons, however flimsy, for authorization of action by the foot soldiers who do things. The bureaucrats don't need to be persuaded; instead, they merely need to be able to point to something that looks plausible enough to justify their actions. Finally, at the top of the pyramid, there are those who formulate, direct, and spread the slopaganda — the sorts of people that Morton calls idealogues. These are the people who *orchestrate* the whole thing. Like the bureaucrats, they may or may not believe what is being propagated or slopagated, but they take advantage of communication tools in order to get their message to an audience that is willing and able to authorize (bureaucrats) and execute (foot soldiers) it. Think Joseph Goebbels, Vladimir Lenin, or Steve Bannon.

This paper unfolds in the following way. First, the phenomenon of slopaganda is characterized and distinguished from other strategies of shaping decision-making at a group level. We pay special attention to the effectiveness and role of GAI in creating and disseminating slopaganda. Second, the effect of slopaganda on individual decision-making is put in the context of what we know about relevant aspects of the cognitive, affective, and sensory capacities of the human brain. The result of this is a hypothesis about the 'difference that makes a difference' (Waters 2007; Scarantino 2015) for slopaganda and a schematic model of factors responsible for its unprecedented effectiveness in shaping group decision-making. Specific examples that fit the model are provided as evidence. Third, the paper offers some actionable responses that *by design* counteract or mitigate the effects of slopaganda.

## 1    Hijacking decision-making of groups by design

The sophistic arts, i.e., the presentation of arguments, selection of support for claims, and effective use of language for persuasion, are important for successful leadership, especially in democratic politics (Kane and Patapan 2010). Political rhetoric, understood to be a skill acquired through learning, has been taught and studied at least since Ancient Greece.[2] Contemporary research in rhetorical psychology (Blumenau and Lauderdale 2024), discourse analysis, and linguistics touch upon the same general phenomenon of persuasive speech in context of contemporary societies (Booth 2009). Of course, the danger of acquiring skills in rhetoric is that they do not always go hand-in-hand with concern for the truth. Rhetoric is not always *mere* rhetoric. It can be used to persuade or reassure people of truths for which there is good warrant or not. When rhetoric is used at scale, it can be more helpfully characterized as propaganda. Here, we do not take an anti-rhetoric or anti-propaganda stance, but rather use these concepts to bring into relief the similarities and differences between them and a new form of persuasion at scale.

---

[2] For example, rhetoric was essential to discharging civic duties in the *boule* (a form of parliamentary council), court, or *agora* (public meeting place).

Methods of propaganda have always been tied to technology. After the Renaissance, the printing press became a tool of mass influence. Pamphlets, gazettes, and posters were the first *by design* means to change the informational ecosystem for large groups sufficiently to facilitate a desired end. After the advent of electrification, new technologies entered the stage. Joseph Goebbels, for example, commissioned the Volksempfänger, or "people's receiver" to broadcast nazi propaganda directly into people's homes. Propaganda is now almost universally associated with totalitarian states, perhaps thanks to George Orwell's *1984* and people like Goebbels. However, propaganda has not always been at the service of totalitarian states (Stanley 2015). The first official institution dedicated to it was in the Vatican, at the service of the Catholic church's evangelizing mission. Propaganda methods were used throughout the church's domains, but especially among Indigenous people of conquered lands administered by European colonizing empires. Napoleonic France repurposed these methods to build a particular image of Napoleon, the Grand Armee, and France. Then, eventually, so did every other major state, including the Soviet Union and the United States. Propaganda is, by definition, propagated: it needs to be spread widely to even count as an instance of the category. Following other scholars, we hold that not all propaganda is false. Even Goebbels famously held that, for the Big Lie to be effective, it had to be intermixed with some truth.

That being said, it is useful to also introduce the notions of misinformation and disinformation in this context. A relatively stable consensus has emerged in the scholarly literature, according to which misinformation is systematically misleading, but not necessarily associated with any intent on the part of the speaker to deceive. By contrast, disinformation is designed to deceive (Aimeur et al. 2023). Propaganda can promote misinformation or disinformation — or neither. Importantly, propaganda can also amplify true information in a way that is calculated to redirect patterns of attention, thereby influencing decision-making without deceit. Roland Barthes (1957) famously discussed examples of this under the heading of *exnomination*. An example of exnomination would be reporting on crime committed by white citizens as "crime" while reporting on crime committed by Black citizens or immigrants as "Black crime" or "immigrant crime," which Breitbart Media has done (Alfano et al. 2018). In recent work, Alfano et al. (2024) showed that GAI (in particular, DALL-E) tends to produce exnominating images when prompted to produce pictures of various religious groups.

Both slopaganda and propaganda aim to influence viewpoints or ideologies, altering the informational environment en masse to reach a desired aggregate change in decision-making at the group level. However, advancements in communication technologies and GAI differentiate them, particularly in terms of the emergence and dissemination of messages in the propaganda narrative. Traditional propaganda relied on media such as yellow press, pamphlets, radio broadcasts, and television. With the rise of the World Wide Web and social media platforms, this evolved into computational propaganda, marked by the use of social bots and algorithmic mechanisms.

Social bots amplify certain narratives while suppressing others by automatically replicating messages across platforms (Woolley & Howard, 2018). Recommender systems further reinforce this process by delivering and exposing users to the same narratives repeatedly (Alfano et al. 2021). This combination of bots and recommendation algorithms defined the era of computational propaganda, characterised by dissemination of content on an unprecedented scale (Fard & Verma, 2022). GAI promises to revolutionise computational propaganda by addressing one of its key limitations. While the latter excelled in message amplification, it lacked personalisation. GAI introduces mass personalisation, creating tailored messages and narratives according to context features or user characteristics. There is emerging evidence that such personalization can be effective at scale and hard to counter with accurate messages (Simchon et al. 2023, 2024; Carella et al. 2025).

We think that slopaganda is qualitatively different from other forms of group influence. First, slopaganda makes new things possible, in particular, to target sub-audiences via market segmentation and individualization. A newspaper appears the same to every reader. A radio broadcast is universal in its message, even if the broadcaster attempts to use dog whistles and figleaves (Saul 2024). Now slopaganda is already being used to "write" many "local" news articles, which are likely to be more trusted than the national or international press. In addition, slopaganda can be targeted to the demographic and psychographic profile of the individual media consumer, based for instance on their social graph, their pattern of previous engagement, their Big Five or Big Six psychological profile, their location, what times of day they tend to be online, and a host of other parameters. Readers will no doubt be familiar with the panic about Cambridge Analytica engaging in micro-targeted political advertising during the 2016 American presidential campaign (Laterza 2021). In our view, this panic was ultimately unjustified, but it was sensitive to a possibility that may now be moving into view with more powerful slopaganda tools enabled by GAI. The efforts of Cambridge Analytica were relatively crude when contrasted with what is now technologically possible. The combination of quantity and quality is what especially concerns us. With elections around the world turning on the knife's edge year after year, even small targeted effects at large scale are dangerous.

Second, slopaganda differs from traditional propaganda on three measurable dimensions by orders of magnitude: scale, scope, and speed (Fox, 2020; Waisbord, 2018). Regarding scale, GAI has made it possible to produce large volumes of fabricated content quickly and at low cost, making it more pervasive and accessible to a wide range of groups. Slopaganda can be produced faster than propaganda produced by any known or foreseeable content-farms. In terms of scope, GAI generates highly personalized and diverse content, allowing messages to be tailored to different audiences and contexts, which means it can generate engagement at scales that other means could only hope for. Finally, the speed of slopaganda leverages the hyper-connectivity of social networks and algorithmic mechanisms, accelerating content dissemination beyond traditional means (Christakis, 2009). On any one of these dimensions, outclassing the competition makes debunking misinformation and disinformation notoriously difficult, as has been noted in discussions of "Brandolini's Law." This "law" was first formulated by Alberto Brandolini in a facetious tweet, stating that "The amount of energy needed to refute bullshit is an order of magnitude bigger than that needed to produce it." The underlying insight has received academic attention and has sometimes also been called the Bullshit Asymmetry Principle (Williamson 2016).

Another less measurable dimension that is affected by slopaganda is persuasiveness, which can be explained through the psychological phenomenon known as the illusory truth effect. Based on this phenomenon, repeated exposure to a statement increases its perceived truthfulness, regardless of its actual validity (Hasher et al, 1977). This phenomenon remains robust to various procedural variables such as subject, statement type, presentation mode, repetition interval, etc. (Dechêne et al, 2010). The illusory truth effect is particularly significant for propaganda, as it offers a mechanism to improve the believability of the propaganda message (Pennycook et al, 2018). The illusory truth effect is so central to propaganda that it underpins the famous law of propaganda often attributed to Goebbels: "Repeat a lie often enough, and it becomes the truth.[3] While various factors contribute to the believability of propaganda (DiFonzo and Bordia, 2007) and mere repetition alone does not account for all of its persuasive power, studies show that familiarity can serve as a cue for truth, thereby boosting a message's believability (Unkelbach et al, 2019). In that sense, the increase in the scale and speed of content production in the slopaganda era may reinforce this effect by repeatedly surfacing the same narratives, which can be seen as an accelerated form of traditional propaganda. However further empirical research is necessary to determine precisely how these high-frequency narratives interact with other psychological and social factors that shape persuasion (Cialdini, 2007).

In summary, slopagada has unique features (targeting), unique magnitudes of features (scale, scope, speed), and unique qualitative improvements (persuasiveness) that together make it distinct from any prior form of group influence strategy. This uniqueness presents unique challenges at the level of the individual and group. Rhetoric and propaganda are by now almost hackneyed and commonplace and consequently may have lost their effectiveness on more sophisticated consumers of information. Slopaganda, however, is new, unexplored, and thus difficult to recognize for what it is even sometimes by the most sophisticated among us. By its effects on individuals, slopaganda poses challenges at a group level, which can only be addressed by new solutions and not those that may have worked with earlier manifestations of political rhetoric and propaganda.

## 2      The cognitive science of slopaganda

### 2.1    Attention Economy

The phrase "attention economy" has become popular in both the scholarly and the popular literature. It was originally coined to refer to the problem that managers of large companies and other institutions faced when dealing with many long reports from their underlings (Simon 1971). It has since come to be associated with a broader sense of information overload felt by everyone with an internet connection. Life is short, and there are only so many hours in the day. At the same time, the amount of available information continues to grow at a pace that no mortal can keep up with. This situation inevitably leads to trade-offs. On what basis should we choose our information sources, given that there are far too many of them? Research suggests that people are especially attracted to sources that both capture and keep their attention (Hofstadter 1964). But what captures and keeps attention? The mechanisms of attention are so general and varied that the taxonomy alone would require a distinct section of this paper (Chun, Golomb, and Turk-Browne 2011). For our purposes here, it suffices to say that emotionally alarming content, threats, and conspiracy theories can all be things that grab attention more than others (Brady et al. 2020).

---

[3] See url = < https://www.bbc.com/future/article/20161026-how-liars-create-the-illusion-of-truth >, accessed 24 February 2025.

This is perhaps unsurprising, given our evolutionary lineage. As finite creatures, it is adaptive for us to be especially attuned to existential threats, even if we score many false positives because of this hyper-attunement. The downstream consequence, however, is that there is a market for threatening messages, as well as a market for rationalizations for responding to such threats in socially harmful ways (Williams 2023). Slopaganda makes it possible to supply an endless stream of plausible threatening messages targeted to individuals and also justifications for them and other threats we may tacitly or explicitly hold at the time. Ultimately, they make it to memory, semantic or long-term, where they can influence individuals over their entire lifetimes or at least long enough to matter for some decision-making.

The relationship between attention and memory is complex and nuanced in a number of ways (van Ede and Nobre 2023). For our purposes, the most important features of attention are its selectivity, relationship to top-down and bottom-up processing, magnitude modulated by arousal/mental state, and temporal grain or scale. The temporal grain of memory typically corresponds to identifiable mechanisms, such as short-term visual memory (Sperling 1960), working memory (Baddelay 1986), or long-term episodic memory (Tulving 1972). These operate at different scales ranging from a few hundred milliseconds to minutes to days and years. Features and mechanisms of attention interact in nuanced ways and in a highly context-dependent way (Cowan et al 2024). In the context of slopaganda, what is important is that 'information overload' creates conditions in which attention selects messages based on features that are not always relevant. This affects what makes it to memory and how it is eventually used to make decisions. In other words, slopaganda can affect attentional mechanisms that gatekeep what makes it to long-term memory relevant to decision-making. With sufficient targeting, this achieves what political rhetoric, propaganda, and mis-/disinformation aim to, but by different (cognitive) means.

## 2.2    Memory, Emotion, and Confirmation-Negativity Biases

Prior knowledge, especially knowledge accumulated by observing the consequences of one's decisions, forms an important basis for future decision-making (Haggard 2008). People learn from their successes and mistakes. People can also learn decision-relevant information without ever having to make decisions. For example, they can learn about the rules that govern driving a car by reading a book about them. This can happen even when this new information is false and maladaptive. For example, the automotive manual can be outdated or contain rules from a different country, or when the consequences of one's actions are misinterpreted or unobserved. What matters to the accumulation of decision-relevant knowledge is that new information is treated as important and relevant enough (Gilboa & Marlatte, 2017; Kemp et al., 2024). As the previous section about attention and memory suggests, what our cognitive and affective mechanisms treat as such (and thereby memory-worthy) is not always important or relevant, ceteris paribus.

For example, we know that when the human brain initially processes new information it involves neural representations that may be similar to and different from other neural representations that constitute our prior knowledge. The magnitude of these similarities and differences, as measured by neuroimaging methods, can influence how well or poorly new information is integrated with prior knowledge (Baldassano et al., 2018; Ezzyat and Davachi, 2021; McClelland et al., 2020; Milivojevic et al., 2016; Schapiro et al., 2013). Neural similarity may be enough to give our brain the signal that what we are hearing, reading, seeing, or experiencing is important enough to integrate with prior knowledge.

Neural similarity can also be neural similarity in unrelated brain events. For example, similar emotional responses or emotions evoked at the time that information is processed can profoundly influence both attention and memory (Cahill & McGaugh, 1995; LeDoux, 1998; McGaugh, 2000). New information interpreted to have emotional significance or information accompanied by emotional states — especially negative ones — is better remembered and integrated with prior knowledge (Kensinger & Corkin, 2003; Rozin & Royzman, 2001).

People remember negative information better, which is sometimes called 'negativity bias.' The most straightforward way this happens with slopaganda is when high negative sentiment leads people to reject information in the general 'slop' presented to them, which is not in line with their own beliefs (Druckman & McGrath, 2019; Kunda, 1990; Nickerson, 1998). The strength of negativity bias also often depends on individual traits (Norris et al., 2019), including personality traits (Calvillo et al., 2024; Vedejova & Cavojova, 2021). So, not all people are affected by emotionally evocative content in the same way, but when we look at the aggregate level of groups, the effect of negative emotions on memory can be pronounced. This is a niche that slopaganda can take advantage of uniquely, since it can target groups based on prior knowledge of what they may react to negatively. For example, people concerned with justice or harm may be disproportionally susceptible to slop that features injustice and depictions of violence; such content alone may be enough to generate the negative sentiment.

Slopaganda also takes advantage of confirmation bias (Nickerson 1998). In short, people actively seek out confirmatory evidence and are more likely to accept information as true when it conforms with their prior knowledge (Kaanders et al 2022). They are less likely to accept information proportionally to the level of conflict with their prior knowledge. With micro-targeted slopaganda, it should be possible, however imperfectly, to estimate an agent's prior beliefs, then serve them content that reinforces pernicious political biases while avoiding showing them alternative perspectives. Likewise, it should be possible, however imperfectly, to target content, for instance to vaccine skeptics with medical misinformation.

Of course, neural representations need to be flexible enough to be updated when already integrated information is discovered to be false or maladaptive. For example, you read a poster that informs you that a famous politician has lied about taking bribes. This is integrated with prior knowledge of corrupt politicians. Suppose you later find out that that politician refused bribes after all. In that case, this politician's name should — both from a self-regarding epistemic perspective and from the moral perspective of avoiding false accusations — be disintegrated from your prior knowledge about corrupt politicians and perhaps integrated with prior knowledge about politicians who refuse bribes. We know this happens routinely.

Correcting information is known to draw attention toward a discrepancy, which then tends to trigger more pronounced confirmation bias (Festinger, 1957; Frey, 1986; Nickerson, 1998; Taber & Lodge, 2006). In other words, when correcting information is in line with what we already believe, confirmation bias makes the integration with prior knowledge stronger than it would have been otherwise. Remembering and interpreting original information (Franks et al., 2023) and corrected information (Walter & Tukachinsky, 2020) is easier when it confirms prior knowledge. The same happens with affect, specifically negative affect, and correcting information. Those whose sentiment is neutral or positive about correcting information simply do not care enough to update their prior knowledge as thoroughly as those who have negative sentiment towards it. Slopaganda can take advantage of these facts through microtargeting, producing content that describes nonexistent threats or exaggerates real threats for people who already harbor problematic beliefs and inclinations. A good example of this may be outrage stories about famous people we love to hate. Finding out that, say, Tiger Woods was never actually smoking crack, it was just marihuana, may not change our model of him as a degenerate drug user.

Finally, and perhaps most importantly for understanding slopaganda, neural representations of information that were shown to be false continue to influence people's beliefs and reasoning after being corrected (Gordon et al., 2019; Sanderson et al., 2023). These erroneous representations do not disappear like the text we delete in a document. A trace of the prior representation typically remains in the brain, influencing how new information is integrated with prior knowledge. We also know that not all people react to being faced with correcting information in the same way. Psychological factors, including beliefs, can determine how new information that corrects errors is accepted and integrated with prior knowledge (Lyons, 2023).

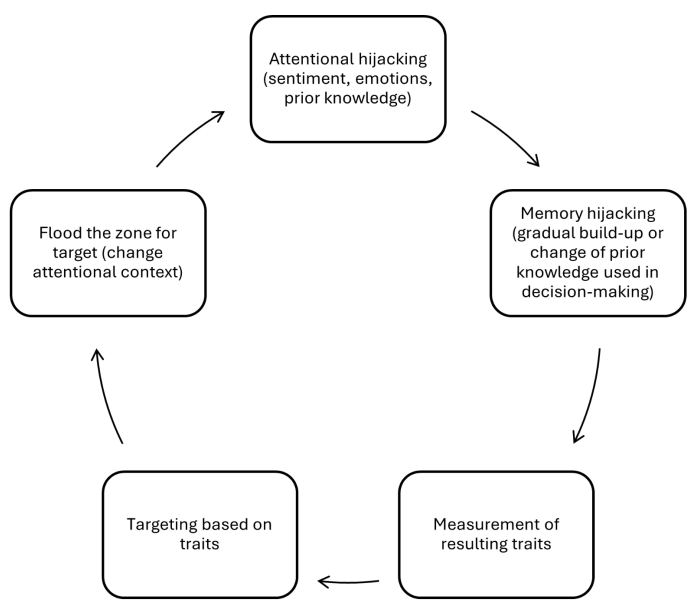

Figure 1. Slopaganda and the Brain Interaction

Putting these two things together, if we know enough about a person, including their psychological traits and prior beliefs, we also know which factors that have nothing to do with misinformation or slopaganda being confronted at the time can determine whether it leaves the relevant trace, even after correction. We can imagine, for example, how knowledge about someone's zip code, how long their commute is, or when they tend to be awake would make them susceptible to entertaining and even accepting claims about the safety and efficacy of vaccines, the advisability of congestion pricing for inner-city commutes, or the desirability of increased immigration. That initial acceptance, even after further information that undermines the claim is accepted, may be enough to lower that person's 'resistance' to future new information along the same lines.

This point is critical because, when people are being influenced at scale, even if effect sizes are tiny, it can sway public opinion, change electoral outcomes, undermine public health efforts, and ensure that people don't all consume safe food products such as pasteurized milk and juice. In a population of hundreds of millions, emotional interventions targeted to people's individual demographics and psychographics are liable to have small but predictable effects. GAI that generates content that people want to hear, whenever they want to hear it, is just the sort of intervention that could serve this function.

## 2.3    Portents of the slopaganda shitstorm

In this section, we review the likely short- and medium-term social consequences of slopaganda. Because we are most familiar with the situation in the United States and because the United States seems to be an epicenter of relevant technological and political developments, we focus on cases from there. To be clear, however, we see no reason to think that slopaganda is or will be limited to the United States in the coming years.

Here is a relatively old example of propaganda that takes advantage of market segmentation, though not individualization and microtargeting, and differences in trust in local versus national media: Sinclair Media's cornering the market in local news broadcasts in the United States nearly a decade ago. A reporter for Deadspin noticed that allegedly local broadcasters were saying oddly similar things to each other, despite residing in distant municipalities. He stitched together video of each of them delivering a warning, which turned out to be identical across all broadcasters.

As it turned out, this was not a massive coincidence. Instead, each of these allegedly local broadcasters was owned by Sinclair, and they had been sent a script for a "forced read," which they could not modify without career repercussions. They dutifully complied. Many watchers of these broadcasts presumably felt a degree of trust in their neighborhood reporters that they might not feel towards more distant national or international figures. As the *New York Times* observed reporting on this:

> The sharing of biased and false news has become all too common on social media.
> Some members of the media use their platforms to push their own personal bias.
> This is extremely dangerous to our democracy.[4]

The fact that this stunt was orchestrated at such a scale is troubling. What is even more troubling is the idea that it can now be amped up by slopaganda that uses not just market segmentation by location but a range of other microtargeting techniques.

Consider another example from the media ecosystem. Steve Bannon, the founder of Breitbart News, bragged that his strategy is to "flood the zone with shit."[5] Media consumers who get their news from Breitbart face a world structured by the categories of Big Government, Big Journalism, Big Hollywood, National Security, Tech, and Sports. This is liable to influence patterns of attention, confirmation bias, and negativity bias. One of the more objectionable semantic tags on stories at Breitbart is 'Black Crime'. (There is no corresponding tag for 'White Crime' or any other racialized crime.) Alfano et al. (2018) found half a dozen stories with this tag, all with sensationalistic headlines, such as, "Black rape gangs violate two Detroit women in one night, hours apart," and "Black mob swarms Georgia Walmart to see 'how much damage' they could do." The consumer of Breitbart will have her experiences structured in a way that may naturally give rise to distinctive biases. Plausibly, a reader of Breitbart Texas, for example, will be led to associate Mexican concepts with crime concepts and negative valence. Such conceptual connections (e.g., Mexico = crime = bad), when ossified, are epistemically problematic because they lead people to make unjustified generalizations and act upon them. This study was published before the rise of GAI. One can imagine what could happen if, instead of half a dozen such stories, the zone truly were flooded with slopaganda shit promoting racial bias.

---

[4] See url = < https://www.nytimes.com/2018/04/02/business/media/sinclair-news-anchors-script.html >, accessed 15 January 2025.

[5] See url = < https://www.vox.com/policy-and-politics/2020/1/16/20991816/impeachment-trial-trump-bannon-misinformation >, accessed 14 January 2025.

In a more recent study, Alfano et al. (2021) examined the YouTube recommender system. They set up a robot to search for videos on a range of topics operationalized by sets of keywords. The robot would then "click" on the top five recommended videos. This process was iterated multiple times to create a branching tree of recommendations. The researchers then manually coded the most recommended videos for each topic based on how conspiratorial they were. Depending on which search terms informed the recommender system, as many as half of the top recommendations promoted conspiracy theories. Importantly, there were differences in how prominent conspiracy recommendations were across topics. Searches for tiny homes and other ways of achieving independence from prevailing socio-economic systems rarely led to conspiratorial content. By contrast, searches for the names of right-wing gurus and influencers such as Jordan Peterson and Ben Shapiro led to conspiratorial content in roughly half of all recommendations. Given the discussion of attention, negativity bias, and confirmation bias above, this makes sense. The sort of person who is most liable to search for these topics likely already embodies morally and politically problematic attitudes. The recommender system gladly reinforced these attitudes. While this example does not involve GAI and therefore does not count as slopaganda, it limns the shape of things to come.

To give just a few more recent examples from US politics: during the 2024 US Presidential campaign, Donald Trump posted GAI slop that suggested that the musician Taylor Swift had endorsed him.[6] He also falsely accused his rival of posting GAI images.[7] Relatedly, Elon Musk posted an AI-enabled deep-fake imitating Kamala Harris's voice, in which her digital impersonator says things that would have been deeply embarrassing and troublesome for the actual candidate to have said.[8] None of these counts as slopaganda, as we define it, because they are one-off acts that involve GAI but rely more on the fact that the actor has a very large audience than on the fact that a large number of allegedly-informative items of prose, image, or video are produced and then distributed in a targeted way. We bring these examples up, however, because we think that it is likely that the slopaganda of the future will employ the tactics of previous generations at greater scale and speed.

To be clear, we also do not think that this problem is isolated to the United States. In 2023, the mayor of a town in Australia was defamed by ChatGPT when a user prompted it to describe the crimes he had committed. He was not in fact a convicted criminal, but ChatGPT obediently accused him of bribery and indicated that he had served prison time. He prepared a defamation lawsuit, which made international headlines, but he subsequently abandoned it.[9] This example was a one-off case prompted by a single user prompting ChatGPT, but again, we are concerned that such content could easily be produced and published at scale and speed, with micro-targeting to make things even worse.

---

[6] See url = < https://www.theguardian.com/technology/article/2024/aug/24/trump-taylor-swift-deepfakes-ai >, accessed 15 January 2025.

[7] See url = < https://www.bbc.com/news/articles/cx2lmm2wwlyo >, accessed 15 January 2025.

[8] See url = < https://apnews.com/article/parody-ad-ai-harris-musk-x-misleading-3a5df582f911a808d34f68b766aa3b8e >, accessed 15 January 2025.

[9] See url = < https://www.smh.com.au/technology/australian-mayor-abandons-world-first-chatgpt-lawsuit-20240209-p5f3nf.html >, accessed 14 June 2025.

# 3 Interventions by design

The remaining question, given the potential and actual consequences of leaving slopaganda to wreak havoc on our informational ecosystems, is whether we can use these insights to counter it *by design*. The short answer is: yes. The longer and more sober answer is that it depends on a number of factors largely outside of our control. These range from the psychological to the technological to the economic to the political. In this concluding section, we canvas each, while remarking on interactions among them.

## 3.1 Psychological interventions

Given that microtargeted interventions relying on psychometric profiling are part of the slopaganda package, it may make sense to counter-target salutary interventions based on people's psychological profiles. In other words, the micro-targeting feature that makes slopaganda unique among influence strategies may also be a key to countering it. For instance, if people who score high or low on some Big Five or Big Six trait (e.g., neuroticism, openness, humility) or a more specific metric (e.g., openmindedness, ingroup criticism, epistemic vice) are especially prone to accepting unwarranted conspiracy theories and medical misinformation (Meyer et al. 2021a, 2021b; Parnamets et al. forthcoming), counter-messaging could be directed towards such people. Research suggests that such counter-messaging is especially effective when it arrives *before* the misinformation or disinformation, in the form of *prebunking* rather than *debunking* (Lewandowsky & Van Der Linden 2021). What is here envisaged is a synchronic psychological intervention that harnesses what we know or can guess about people. As researchers in misinformation and disinformation use the term 'prebunk', it refers to warning people in advance that they are likely to encounter misleading messages, sometimes explaining the motives and tactics of those who purvey such messages. Prebunking is typically contrasted with debunking, which occurs after people have already encountered the misleading messages. There is evidence that prebunking is more effective than debunking (see Lewandowsky & Van Der Linden 2021 and the literature reviewed therein).

Nudging is another psychological intervention that influences behaviour while preserving individuals' freedom of choice (Thaler & Sunstein, 2008). Various nudge-based strategies have been developed to reduce the spread of false or unverified information. For instance, accuracy nudges prompt users to evaluate the veracity of information before sharing it, thereby encouraging the dissemination of reliable content (Pennycook & Rand, 2022). Social-norms nudges leverage people's tendency to conform to community standards by subtly reminding individuals of widely accepted behaviours, which promotes more accurate information sharing (Butler et al., 2024). Other strategies include information nudges that provide contextual information such as labels that indicate a source's reputability or political leaning to raise awareness about potential biases, and presentation nudges that frame choices through the structured placement of content (Thornhill et al., 2019). Additionally, prosocial nudging engages individuals on a personal level: for example, asking participants to write a letter to a less digitally competent relative reframes misinformation detection as an act of helping loved ones, thereby increasing personal investment in avoiding fake news (Orosz et al., 2023).

Games serve as another psychological intervention against misinformation by employing inoculation techniques that expose players to a variety of deceptive tactics. Titles such as Fake It To Make It (Urban et al., 2018), Bad News (Roozenbeek & Van der Linden, 2019), Harmony Square (Roozenbeek & van der Linden, 2020), Go Viral! (Basol et al., 2021), and Cranky Uncle (Cook et al., 2023) immerse users in interactive environments where various misinformation tactics, from emotional manipulation and polarisation to conspiracy-building and climate-related distortions are clearly shown. By taking on roles like misinformation entrepreneurs or fake news creators, players gain insight into the mechanics behind deceptive information, which helps to improve media literacy and critical thinking in digital contexts.

Preliminary empirical studies based on data collected from these games offer support for the effectiveness of such interventions. For example, research on Bad News, Harmony Square, and Go Viral! shows that exposing individuals to weakened doses of misinformation can influence their ability to spot and resist misleading narratives (Axelsson et al 2024). Similarly, the Cranky Uncle game demonstrates that engaging gameplay with humour can reduce the perceived reliability of false information and lower the chance of its spread. However, these findings should be interpreted with caution, as more systematic research is needed to establish their effectiveness and long-term impact.

More diachronic interventions would target people over time. We see two as especially promising. First, educating for digital literacy will be an absolute necessity in the coming years and decades (Guess et al., 2020). We are not the first to advocate for digital literacy (Diepeveen & Pinet 2022), but we think that it is especially pertinent in the current era, given the swift and unregulated strides that GAI has taken and the potential for malfeasance by those who might deploy slopaganda. Second, there are more time-intensive and effortful ways to cultivate the dispositions (cognitive, affective, or intellectual) that may immunize people against propaganda and slopaganda to some degree. To date, the most promising way to increase intellectual humility involves self-distancing in journal-writing over the course of one month (Grossman et al. 2021)—that is, referring to oneself using a third-person pronoun or one's name, rather than the first-person singular. This is a high-cost intervention that may not generalize to other contexts. That being said, it could be piloted as a potential further curb on slopaganda.

## 3.2    Technological interventions

The psychological and educational interventions canvassed above are welcome, but we know from decades of research that such interventions produce only small effects at great cost (Grossman et al. 2021). For this reason, we think that additional vectors of intervention against slopaganda may be warranted. These could be developed by university researchers, collaborations between universities and industry, collaborations between universities and governments, or any combination of the above. It is imperative that these interventions be seen and received as benevolent offers rather than attempts to censor or contravene the inquiries of various publics. The latter perception would likely lead to a backfire effect, or at least undermine the effectiveness of any interventions.[10] Instead, we suggest meeting people where they are and helping those who want to overcome their own prejudices and confirmation bias to do so in a transparent way.

---

[10] The backfire effect has come under serious scrutiny in recent years (Nyhan 2021). We do not take a stand on the the effect as initially described, but we find it highly plausible that receiving a debunking or prebunking message from a source that is viewed as untrustworthy is not liable to improve someone's epistemic condition.

One of the technological interventions to counter harmful content on online platforms is content moderation. This process enforces community guidelines by detecting and managing content that violates rules (Horta Ribeiro et al., 2023). Typically, the process includes setting community standards, identifying violations, enforcing rules, and disclosing moderation activities (Clune & McDaid, 2024). Over time, content moderation has evolved from manual human review to automated systems, then to hybrid approaches, and most recently to crowd-based community notes (Gongane et al., 2022; Singhal et al., 2023; Chuai et al., 2024).

Automated moderation employs AI, often using matching algorithms or classifiers, to handle the vast volume of social media content. This approach offers speed, scalability, and consistency, enabling near real-time intervention. However, it remains susceptible to issues such as over-moderation, under-moderation, contextual misunderstandings, and algorithmic bias (Gorwa et al., 2020). In contrast, human moderators excel at interpreting context but struggle with scalability and consistency, and their well-being can be compromised by exposure to harmful content. The hybrid model benefits from the strengths of both systems by having AI flag potential issues for human review; however, challenges such as evasion tactics, bias, and the lack of AI transparency persist (Gongane et al., 2022; Sheng, 2022; Clune & McDaid, 2024; Singhal et al., 2023). Different platforms continuously update their policies to address emerging challenges such as misinformation and hate speech (Singhal et al., 2023; Schaffner et al., 2024). Once content is flagged and determined to violate community guidelines, enforcement actions can range from content removal, content hiding, issuing warnings, content demotion (to reduce visibility), suspending or banning accounts, and labelling (Horta Ribeiro et al., 2023; Gongane et al., 2022; Fard & Lingeswaran, 2020).

Research on the overall effectiveness of content moderation is inconclusive. While some studies suggest that content moderation can positively influence user behaviour by encouraging users to follow the platform rules over time (Horta Ribeiro al., 2023), others argue that these systems are less effective in promoting a culture of responsible content creation, primarily because they often rely on punitive measures rather than on engagement, education, and transparent dialogue (Clune & McDaid, 2024).

We also envisage, for instance, an intellectual humility plugin that encourages self-reflection while scrolling one's social media newsfeed (especially when controversial keywords and hashtags occur in posts), discourages reposting without having clicked through, and makes recommendations for both follows and unfollows that would connect users with a larger variety of diverse, independent, trustworthy sources while disconnecting them from homogenous, densely-interconnected, untrustworthy sources. Such a plugin does not, to our knowledge, currently exist, but publicly-minded academics or technologists could at least build a prototype based on existing research.

These technological interventions are only likely to succeed if they are released in a salubrious environment, which the current state of global affairs does not seem to embody. For this reason, we think that more fundamental interventions may be warranted. The solution to epistemic problems such as misinformation, disinformation, propaganda, and slopaganda may not themselves be primarily epistemic. Instead, what may be needed are interventions at the level of economic and political institutions that have been corroded and corrupted over the course of decades.

### 3.3 Politico-economic interventions

The final, and perhaps most important, vector of intervention is politico-economic. Slopaganda will no doubt be produced and distributed for many reasons, from personal to tribal to corporate to state-based. We are especially concerned about the latter two because of the enormous power that large multinational corporations and empires possess. In T*he Wealth of Nations*, Adam Smith (1776 / 1999) devoted part of the fourth book of his magnum opus to a withering critique of the British East India Company and the Dutch East India Company, arguing that "The government of an exclusive company of merchants is, perhaps, the worst of all governments for any country whatever" because such a company occupies a hybrid role of both merchant and sovereign. Such an organization is, according to Smith "a strange absurdity" in corporate political economy. "As sovereigns, their interest is exactly… that of the country they govern [e.g., India or Indonesia]. As merchants, their interest is directly opposite to that interest." The Dutch East India Company, which infiltrated the state after going bankrupt in 1799, was perhaps the worse offender of the two. In current times, OpenAI and related "Magnificent Seven" companies (Apples, Microsoft, Amazon, Alphabet, Meta, Nvidia, and Tesla) have assumed similar roles in national and international economies, facilitating and directing the legislation, policy-making, and judicial systems of entire countries. The most egregious recent example is Elon Musk's capture of the United States bureaucracy via his unelected and unappointed Department of Government Efficiency (trollingly and self-indulgently nicknamed DOGE).

As mentioned above, Musk and Trump have been front-of-stage in the spreading of slopaganda and related forms of misinformation and disinformation. To curb these efforts, educating university students is obviously inadequate. It's far too little, far too late. In this concluding section, we want to suggest that the solution to the current social epistemic problem of slopaganda may not itself be social epistemic, at least in the first instance. The problem here is material. In particular, it is political and economic because it has to do with who has power over whom. The revival of oligarchy was already being documented by Picketty (2014) in the wake of the Great Recession of 2008-9. Beyond a certain level of income- and wealth-accumulation, money has decreasing marginal value in exchange terms (i.e., when being used in exchange for goods or services). The small global elite that controls the majority of generational and new wealth cannot possibly hope to spend that money in a way that directly contributes to their welfare or the welfare of their families and friends.[11] Instead, many of them seem intent on deploying their wealth to interfere in domestic and international politics, including through the distribution of slopaganda. Democracies around the world suffer from these narcissistic efforts to pollute the epistemic commons, as do citizens, denizens, immigrants, and refugees. We therefore conclude with the bold suggestion that one of the most promising interventions to counter slopaganda and related problems is a global wealth tax that would simultaneously reduce the power of oligarchs and fund the sorts of interventions outlined above. This suggestion may strike readers as infeasible, but we think that now is the time to consider bold, infeasible proposals, as infeasibility is often cynically used as an objection by those who benefit unjustly from the status quo (Southwood 2016; 2018). If this bold suggestion is deemed a bridge too far, we nevertheless contend that other political and economic means should be considered to curb the malign influence of slopaganda.

---

[11] See, for example, url = < https://oxfamilibrary.openrepository.com/bitstream/handle/10546/621477/bp-survival-of-the-richest-160123-summ-en.pdf >, accessed 15 February 2025.

*Michał Klincewicz, Tilburg School of Humanities and Digital Sciences, TSHD: Department of Cognitive Science and Artificial Intelligence*
*Mark Alfano, School of Humanities, Discipline of Philosophy, Macquarie University*
*Amir Ebrahimi Fard, Independent Scholar*